\documentclass[draft]{mn2e}
\title[Spitzer's view of Sakurai's Object]
      {The Spitzer IRS view of V4334~Sgr (Sakurai's Object)}
\author[A. Evans et al.]
      {A. Evans$^{1}$, V. H. Tyne$^{1}$, J. Th. van Loon$^{1}$, B. Smalley$^{1}$,
      T. R. Geballe$^{2}$,\newauthor
      R. D. Gehrz$^{3}$, C. E. Woodward$^{3}$, A. A. Zijlstra$^{4}$, E.
      Polomski$^{3}$, M. T. Rushton$^{1}$, \newauthor 
      S. P. S. Eyres$^{5}$, 
      S. G. Starrfield$^{6}$, J. Krautter$^{7}$, R. M.
      Wagner$^{8}$  \newauthor
      \mbox{}\\            
      $^{1}$Astrophysics Group, Keele University, Keele, Staffordshire, ST5 5BG,
      UK \\
      $^{2}$Gemini Observatory, 670 N. A'ohoku Place, Hilo, HI\,96720, USA \\
      $^{3}$Department of Astronomy, School of Physics \& Astronomy,
      116 Church Street S.E., University of Minnesota, \\
      \mbox{~~} Minneapolis, MN 55455, USA\\      
      $^{4}$Department of Physics \& Astronomy, University of Manchester,
      P. O. Box 88, Manchester, M60 1QD, UK   \\
      $^{5}$Centre for Astrophysics, University of Central Lancashire,
      Preston, PR1 2HE, UK \\
      $^{6}$Department of Physics \& Astronomy, Arizona State University,
      Tempe, AZ 85287, USA \\
      $^{7}$Landessternwarte, K\"{o}nigstuhl, D-69117 Heidelberg, Germany \\
      $^{8}$Large Binocular Telescope Observatory, University of Arizona,
      933 North Cherry Avenue, Tucson, AZ 85721, USA
}

\date{Revised version}

\pagerange{\pageref{firstpage}--\pageref{lastpage}}

\def\LaTeX{L\kern-.36em\raise.3ex\hbox{a}\kern-.15em
    T\kern-.1667em\lower.7ex\hbox{E}\kern-.125emX}

\newcommand{\Mdot}[2]{\mbox{${#1}\times10^{-{#2}}$\,M$_\odot$~yr$^{-1}$}}
\newcommand{\Msun}{\mbox{\,M$_\odot$}}
\newcommand{\Lsun}{\mbox{\,L$_\odot$}}
\newcommand{\vunit}{\mbox{\,km\,s$^{-1}$}}
\newcommand{\mic}{\mbox{$\,\mu$m}}
\newcommand{\pion}[2]{{#1}\,{\sc {#2}}}
\newcommand{\fion}[2]{[{#1}\,{\sc {#2}}]}
\newcommand{\ltsimeq}{\raisebox{-0.6ex}{$\,\stackrel 
	{\raisebox{-.2ex}{$\textstyle <$}}{\sim}\,$}} 
\newcommand{\gtsimeq}{\raisebox{-0.6ex}{$\,\stackrel
	{\raisebox{-.2ex}{$\textstyle >$}}{\sim}\,$}}
\newcommand{\prsimeq}{\raisebox{-0.6ex}{$\,\stackrel 
	{\raisebox{-.2ex}{$\propto$}}{\sim}\,$}} 
\newcommand{\acet}{\mbox{C$_2$H$_2$}}
\newcommand{\diacet}{\mbox{HC$_4$H}}
\newcommand{\triacet}{\mbox{HC$_6$H}}

\begin{document}
\label{firstpage}
\maketitle

\begin{abstract}
We present an observation of the very late thermal pulse object 
V4334~Sgr (Sakurai's Object) with the Infrared Spectrometer (IRS)
on the Spitzer Space Telescope. The emission from 5-38\mic\ is
dominated by the still-cooling dust shell. A number of features
are seen in absorption against the dust shell, which we attribute
to HCN and polyyne molecules. We use these features to determine
the $^{12}$C/$^{13}$C ratio for the absorbing gas to be
$\sim3.2^{+3.2}_{-1.6}$; this implies that, despite the H-content
of the molecules, the hydrocarbon-bearing gas must have originated
in material produced in the very late thermal pulse. We see no
evidence of emission lines, despite the recently-reported optical
and radio observations that suggest the effective temperature of the
stellar remnant is rising.
\end{abstract}


\section{Introduction}

\begin{figure*}
\setlength{\unitlength}{1cm}
\begin{center}
\leavevmode
\begin{picture}(5.0,4.5)
\put(0.0,4.0){\includegraphics{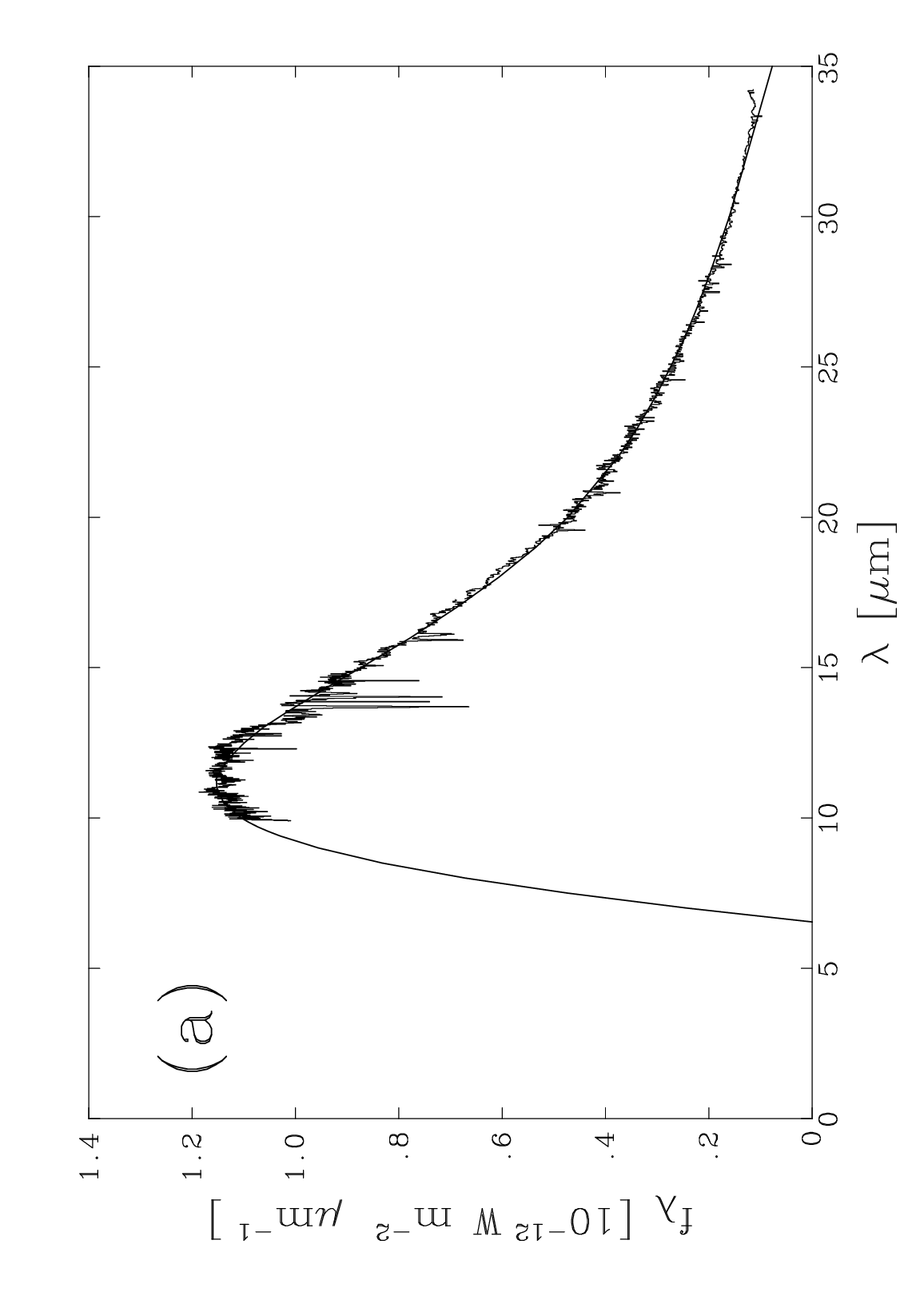}}
\put(0.0,4.0){\includegraphics{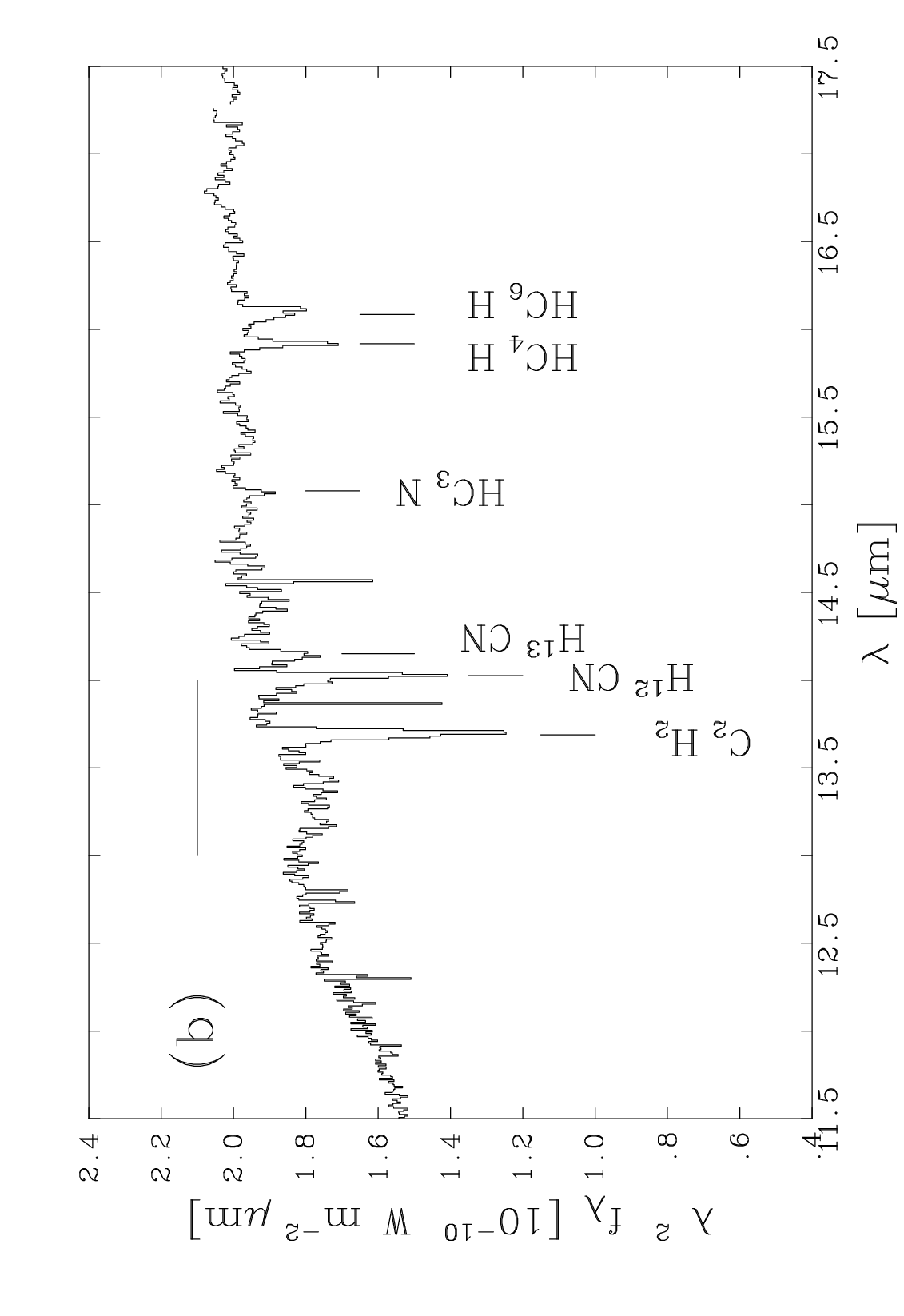}}
\end{picture}
\caption[]{{(a)}: The Spitzer IRS spectrum of Sakurai's Object in 2005.
The curve is a {\sc dusty} fit to the data; see text for
details. {(b):} Detail in the spectrum in the wavelength range
12.5-16.5\mic, plotted as $\lambda^2\,f_\lambda$ to emphasize the depression
(which is indicated by horizontal line) around 13\mic; much of the scatter is
real, arising in ro-vibrational bands of the identified molecules.}
\end{center}
\label{SED}
\end{figure*}

V4334~Sgr (Sakurai's Object) is a low-mass ($\sim0.5$\Msun) star
that is retracing its post-Asymptotic Giant Branch (AGB) evolution
along the Hertzsprung-Russell diagram following a very late thermal
pulse (VLTP) \cite{herwig}. The star was discovered in early 1996
\cite{nakano}, and was found to be at the centre of a faint
planetary nebula (PN) of diameter $40''$ (e.g. Pollacco 1999).
It first appeared as an F supergiant, possibly with a hot dust shell
\cite{duerbeck96}.

Analysis of the optical spectrum for mid-1996 by Asplund et al.
\shortcite{asplund99} revealed that Sakurai's Object had undergone a
transformation to a hydrogen-deficient, carbon-rich star and that the
elemental abundances of a range of species were close to those found
in the R~CrB stars. They also used the C$_2$ (1-0) and (0-1) Swan bands
at 4740\AA\ and 5635\AA\ respectively to determine that the
$^{12}$C/$^{13}$C ratio was in the range 1.5-5. Pavlenko et al.
\shortcite{pavlenko04} used the first-overtone CO 
bands to determine that the $^{12}$C/$^{13}$C ratio was $4\pm1$.
This is substantially less than the value in (for example) the Solar
System ($^{12}$C/$^{13}$C = 70; Mathis 2000) and likely arises from
second stage CNO cycling following He burning in the VLTP \cite{asplund99}.

In 1998, Sakurai's Object was completely obscured by an optically thick
dust shell from which (as of December 2005) it has still not emerged. Indeed
analysis of the infrared (IR) spectrum showed that dust emission has been
making an increasing contribution to the IR emission since 1997 April
\cite{pavlenko02}. Since the major dust event of 1998, the 1-5\mic\
spectrum is black body-like and is that of the ejected dust shell.
Analysis of the post-1998 IR observations has shown that the dust is
carbon, primarily in amorphous form \cite{tyne1,eyres}. The 1-5\mic\
spectra during the period 1999-2001 were consistent with massloss
that increased from $\sim\Mdot{0.5}{5}$ to $\sim\Mdot{1.2}{5}$ for a
gas-to-dust ratio 200, appropriate for solar abundances (note that a
lower gas-to-dust ratio of 75 was suggested by Evans et al.
\shortcite{evans-jcmt} for the hydrogen-deficient,
carbon-rich wind of Sakurai's Object). The maximum grain radius (in a
grain size distribution $n(a)\,da \propto a^{-q}\,da$, with $q\simeq3$)
increased by a factor $\sim3$ \cite{tyne1}.

Sub-millimetre observations \cite{evans-jcmt} show that the flux density at
450\mic\ and 850\mic\ was still rising in 2003, indicating that the
massloss was being maintained; the spectral energy distribution (SED)
resembled a black-body at 360~K.

Observations of \fion{N}{ii}$\lambda6548,6583$ and \fion{O}{ii}$\lambda7325$
in the spectrum of Sakurai's Object in 2002 \cite{kerber} hinted that the
stellar temperature had increased from $\sim5\,500$~K in 1997 to
$\gtsimeq20\,000$~K. The reheating seems to be confirmed by recent
observations at 8.6~GHz with the VLA \cite{hajduk}, which detected
free-free emission from the newly-ionized gas.

We describe here observations of Sakurai's Object with the Infrared
Spectrometer (IRS) \cite{houck} on the Spitzer Space Telescope \cite{werner}.

\vspace{-5mm}

\section{Observations}

Sakurai's Object was observed in STARE mode with the IRS on 2005 April 15 UT
(PID~03362; AOR Key 10840320). Spectra were obtained with both low
and high resolution modes, covering the spectral range of 5--38\mic, and
the RED peak-ap array used to precisely centre the object in the IRS
slits. We executed five 6-second ramps for a total observation
time (including overheads) of 1042.6~s. The spectra in the
wavelength range $\sim5.1-8.2$\mic\ were saturated and are not
presented here. The spectrum was extracted from
the version 12.3 processed pipeline data product using {\sc spice} version
1.1 \cite{spice}. High resolution segments were stitched together
without any continuum normalization.

\vspace{-4mm}

\section{Discussion}

\subsection{The continuum}
\label{cont}

The dust shell, which condensed and became optically thick in 1998
\cite{duerbeck02} still dominates the IR SED from 5--38\mic\
(see Fig.~\ref{SED}{a}). Much of the scatter in the data in
Fig.~\ref{SED}{a} consists of low-level features superimposed on
the dust continuum (see Fig.~\ref{SED}{b}), and will be discussed
in detail in a future paper. The depression in the SED around 13\mic\
likely is due to the superposition of many ro-vibrational bands of
hydrocarbon molecules (see \S\ref{absorption} below).

\begin{table*}
\begin{center}
\caption{Dust shell parameters for Sakurai's Object. $T_*$ is the stellar
temperature assumed for the {\sc dusty} fit, $r_1$ is the inner radius of the
dust shell, $R_*$ is the radius of the star, $\tau_V$ is the visual optical
depth through the dust shell, $f$ is the graphitic carbon fraction, $a_{\rm
min}$ and $a_{\rm max}$ are respectively the minimum and maximum grain radius, 
$q$ is the exponent in the grain size distribution and $T_1$ is the dust
temperature at $r_1$.}
\begin{tabular}{ccccccccccc} \hline
Facility & UT Date     & $T_*$ & $r_1$         & $r_1/R_*$ &$\tau_{\rm V}$  & $f$           & $a_{\rm min}$  & $a_{\rm max}$  & $q$  & $T_1$ \\
         &             & (K)   & ($10^{14}$cm) &                 &                &               &($\!\mic$)      & ($\!\mic$)     &      & (K) \\ \hline
UKIRT    & 2001-Sep-08 & 5\,200    &  $3.33\pm1.2$  & $57.9\pm3$ & $8.9\pm0.3$      & $0.80\pm0.06$ & 0.012                    &     2.84            & $3.07\pm0.09$ & $854\pm25$ \\
Spitzer  & 2005-Apr-15 & 35\,000   &  $25.7\pm0.5$ & $13.4\times10^3$ & $8.8\pm0.2$       & $0.54\pm0.02$   & 0.005   &     2.00            & $2.9\pm0.1$      &$407\pm 10$\\
         &             &           &             & \multicolumn{1}{l}{$\pm0.3$}         &                &    &   &      &      &\\ \hline
\end{tabular}
\label{dusty}
\end{center}
\end{table*}

\begin{figure}
\setlength{\unitlength}{1cm}
\begin{center}
\leavevmode
\begin{picture}(5.0,4.)
\put(0.0,4.0){\includegraphics{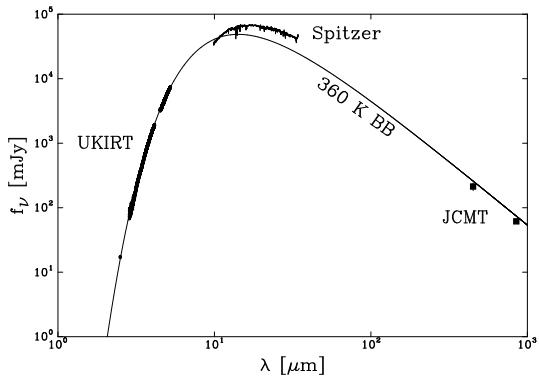}}	 
\end{picture}
\end{center}
\caption[]{The cooling of the dust shell since 2003. Solid curve labelled
``360~K BB'' is a fit to UKIRT (squares in the range 1-5\mic) and JCMT
(squares at 450 and 850\mic) data obtained in 2003 \cite{evans-jcmt};
data labelled ``Spitzer'' are reported here.}
\label{jcmt}
\end{figure}

We compare the Spitzer IRS spectrum with data obtained with the United
Kingdom Infrared Telescope (UKIRT) and the James Clerk Maxwell Telescope
(JCMT) (from Evans et al. 2004) in Fig.~\ref{jcmt}. Compared with the
2003 SED the peak emission has clearly shifted to longer wavelengths and
the peak flux density has increased in 2005. The dust shell superficially
resembles a blackbody with temperature $T$ $\ltsimeq200$~K, significantly
cooler than in 2003 \cite{evans-jcmt}, when the ``blackbody''
temperature was 360~K. The flux density $f_\lambda$ peaks at
$\lambda_{\rm max} \simeq 11.3$\mic, implying a temperature for the dust
``photosphere'' of $\sim210$~K if the (carbon) dust is amorphous with
$\beta\simeq1$ (where the $\beta$-index is defined such that the
emissivity is $\propto\nu^\beta$), or is $\sim180$~K if the dust is
graphitic ($\beta\simeq2$).

The integrated flux is $1.77\times10^{-11}$~W~m$^{-2}$, yielding a
bolometric luminosity $L_* = 2\times10^3$\Lsun\ for a distance of 1.9~kpc
\cite{kimeswenger1}. As the dust shell is still optically thick at
short wavelengths, this radiation is effectively reprocessed stellar
radiation and, assuming that there are no `holes' in the dust
distribution (see \S\ref{hcn}), a measure of the star's bolometric
luminosity. This is somewhat lower than the last reported $L_*$ in
Tyne et al. \shortcite{tyne1}, who found $L_*=4.47\times10^3$\Lsun.
The modest decline in $L_*$, accompanied by the reported rise in
$T_*$, is consistent with the evolution of Sakurai's Object as
predicted by Hajduk et al. \shortcite{hajduk}.

\begin{figure}
\setlength{\unitlength}{1cm}
\begin{center}
\leavevmode
\begin{picture}(5.0,4.)
\put(0.0,0){\includegraphics{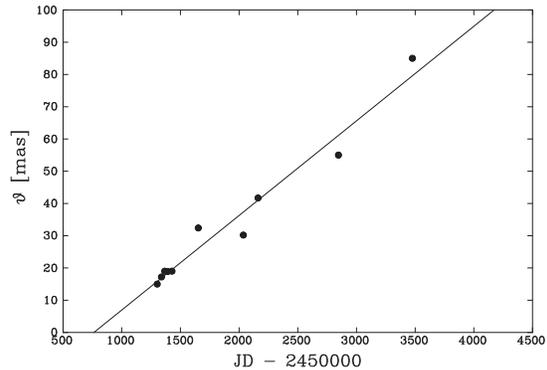}}	  
\end{picture}
\end{center}
\caption[]{Variation of blackbody angular diameter with time;
see text for details.} 
\label{bbang}
\end{figure}

From the IRS spectrum, the ``blackbody'' angular diameter \cite{gallagher}
was 85~mas at the time of our observation, while that  deduced from
the 2003 UKIRT/JCMT data is 55~mas. In Fig.~\ref{bbang} we plot the blackbody
angular diameter as a function of time; earlier values are taken from the data
of Tyne et al. \shortcite{tyne1}.
The angular diameter data are consistent with linear expansion of the dust
photosphere at the rate 0.0293~mas~day$^{-1}$, with the expansion starting
at JD~2450763 (1997 November 10). Comparison with the visual light curve in
Tyne et al. \shortcite{tyne1} shows that this epoch coincides closely with
the time when the light curve began its deep minimum (between JD~2450760
and JD~2450850, i.e. between 1997 November 7 and 1998 February 5).

We fit the Spitzer dust SED using the {\sc dusty} code \cite{dusty} with
the downhill simplex routine we used to fit the 1--5\mic\ data for
Sakurai's Object \cite{tyne1}, assuming a stellar temperature of
35\,000~K \cite{kerber,hajduk}. We solve for visual optical depth ($\tau_{\rm
V}$), graphitic carbon fraction $f$, maximum ($a_{\rm max}$) and
minimum ($a_{\rm min}$) grain size in a distribution $n(a)\,da \propto
a^{-q}\,da$, the exponent $q$, the temperature $T_1$ at the inner radius
$r_1$ of the dust shell and the ratio $r_1/R_*$, where $R_{*}$ is
the stellar radius. The fit is included in Fig.~\ref{SED} and the
derived parameters are listed in Table~\ref{dusty}. For comparison,
Table~\ref{dusty} also includes the corresponding parameters for the
{\sc dusty} fit to the 2001 September 8 UKIRT data \cite{tyne1}.

There have clearly been substantial changes in the values of $f$, $T_1$
and $a_{\rm max}$ since 2001. While the change in $f$ implies conversion
of graphitic to amorphous carbon we should not, as noted by Tyne et al
\shortcite{tyne1}, interpret this change too literally as {\sc dusty}
computes average optical constants for a grain mix. The decrease in
$a_{\rm max}$ (Table~\ref{dusty}) does however support our contention that
a significant change in the character of both the dust mix and distribution
has occurred since 2001. Some processing of the dust has undoubtedly taken
place, possibly a result of the grains' exposure to the hardening radiation
field \cite{kerber,hajduk}; similar changes are seen in the newly-condensed
carbon dust around erupting novae \cite{gehrz98}.

Given the observed increase in the effective temperature of the star
\cite{kerber,hajduk}, the change in $T_1$ may imply that the inner
boundary of the dust shell is receding from the stellar remnant, and
that the condensation of new dust has ceased. If
$L_*$ were constant, then $r_1 \prsimeq T_1^{-2}$ implying
(for no change in dust absorptivity) a $\sim4.5$-fold increase in $r_1$
since 2001. Note also the large increase in $r_1/R_*$, in line with the
rise in $r_1$ and the decline in $R_*$ as the effective temperature of
the star increases.

The increase in $r_1$ is strong indication that the massloss rate (as
measured by the dust) has declined since 2001. If this is
the case then it augurs well for the dispersal of the dust and the
imminent reappearance of the stellar remnant as the dust disperses.
However, in view of the way in which {\sc dusty} treats optical constants
in a grain mix we again caution against over-interpretation of the results,
other than to note the persistent decrease in $T_1$, and the persistent
rise in $r_1$, since late 1997.

\subsection{Spectral features}

\begin{table}
\begin{center}
\caption{Absorption features in Sakurai's Object; identifications from Aoki et al. (1999) and
Cernicharo et al. (2001).}
\begin{tabular}{cccc} \hline
$\lambda_{\rm obs}$ ($\!\mic$) & ID & Band & $\lambda_{\rm id}$ ($\!\mic$) \\ \hline
13.688 & \acet\ & $\nu^5$ $Q$ branch & \\
14.025 & H$^{12}$CN    & $2\nu_2^2-1\nu_2^1$ & 14.00    \\
       & H$^{12}$CN    & $1\nu_2^1-0\nu_2^0$ & 14.04    \\
14.135 & H$^{13}$CN    &   Q                 & 14.1605  \\
15.079 & HC$_3$N       &   $\nu_5$           &          \\
15.918 & \diacet       & $\nu_8$             &  15.926  \\
16.085 & \triacet      & $\nu_8$             & 16.067   \\
           & \triacet  & $\nu_{11}$          & 16.095   \\
\hline
\end{tabular}
\label{ID}
\end{center}
\end{table}

\subsubsection{Absorption features}
\label{absorption}
The SED is far from smooth over the entire IRS wavelength range. There are
several obvious absorption features superimposed on the dust continuum, at
13.688\mic, 14.025\mic, 14.135\mic, 15.079\mic, 15.918\mic, 16.085\mic\
(see Fig.~\ref{SED}{b}). The only plausible identifications for these
features (see Table~\ref{ID}) are H$^{12}$CN, H$^{13}$CN, HC$_3$N, and the
carbon chain molecules (polyynes) acetylene (\acet), diacetylene (\diacet)
and triacetylene (\triacet).

We use the HCN features to estimate the columns of H$^{12}$CN and H$^{13}$CN,
and temperature of the absorbing material. For H$^{12}$CN we take Einstein
coefficients and other relevant data from Harris, Polyansky \& Tennyson
\shortcite{harris}, and partition functions from Barber, Harris \& Tennyson
\shortcite{barber}; the corresponding quantities for H$^{13}$CN
were kindly provided by Dr G. Harris (private communication). The optical depth
in the HCN features was determined relative to a local continuum in the
13--17\mic\ wavelength region.

\begin{figure}
\setlength{\unitlength}{1cm}
\begin{center}
\leavevmode
\begin{picture}(5.0,4.)
\put(0.0,4.0){\includegraphics{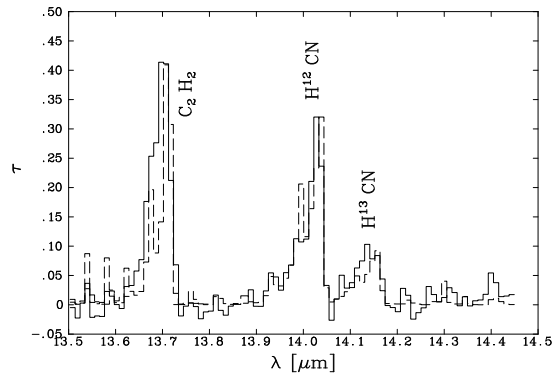}}	 
\end{picture}
\end{center}
\caption[]{Fit to the HCN and \acet\ features. Solid
line, observed optical depth in HCN and \acet\ lines; broken line,
optical depth for column density $N=1.42\times10^{17}$~cm$^{-2}$ (H$^{12}$CN),
$4.51\times10^{16}$~cm$^{-2}$ (H$^{13}$CN) and $9.0\times10^{17}$~cm$^{-2}$
(\acet), and gas temperature 450~K. See text for details.}
\label{HCN}
\end{figure}

We consider an elementary model in which the absorbing gas is homogeneous and
isothermal and located in front of the dust shell; each HCN transition is
assumed to have intrinsic width 10~cm$^{-1}$ ($\sim2\times10^{-5}$\mic) and
broadened by a gaussian having width corresponding to the resolution of
the IRS. The column density-temperature ($N,T$) parameter space was searched,
independently for H$^{12}$CN and
H$^{13}$CN, to find the best fit; we find (see Fig.~\ref{HCN})
\begin{enumerate}
\itemsep=2mm
\itemindent=-2mm
\item [{}] $N = 1.18^{+0.53}_{-0.35} \times 10^{17}$~cm$^{-2}$, $T=400\pm100$~K
(H$^{12}$CN) and
\item [{}] $N = 5.35^{+2.06}_{-1.55} \times 10^{16}$~cm$^{-2}$, $T=500\pm100$~K
(H$^{13}$CN).
\end{enumerate}
The greatest uncertainty is in the column rather than temperature; the line
profile is sensitive to the latter, while the former is more sensitive to the
placement of the continuum. We do not expect the temperatures of
the H$^{12}$CN and H$^{13}$CN regions to differ, and the above values of
$T$ are consistent within the errors. For the purpose of discussion we
take $T=450$~K, for which $N=1.42\times10^{17}$~cm$^{-2}$ 
(H$^{12}$CN) and $N=4.51\times10^{16}$~cm$^{-2}$ (H$^{13}$CN; see
Fig.~\ref{HCN}). This immediately gives us the $^{12}$C/$^{13}$C ratio as
$\sim3.2^{+3.2}_{-1.6}$, consistent with the
value $4\pm1$ obtained by Pavlenko et al. \shortcite{pavlenko04}
from fitting the first overtone CO bands in the near-IR, and
with the VLTP interpretation of Sakurai's Object. 

We note that the $\nu_5$ band of $^{13}$C$^{12}$CH$_2$ is blended with the
corresponding $\nu_5$ band of $^{12}$C$_2$H$_2$ \cite{cernicharoetal99}
and, at the resolution of the Spitzer IRS, can not be used to provide an
alternative estimate of the $^{12}$C/$^{13}$C ratio. Similarly, we do not
expect to see the effect of $^{12}$C$\:\:\rightarrow\!^{13}$C substitution
in the polyynes in our data. However, for $T=450$~K and assuming the
$^{12}$C/$^{13}$C ratio deduced from the HCN lines, the column density
for \acet\ is $9.0\times10^{17}$~cm$^{-2}$, where we have taken partition
functions from Gamache, Hawkins \& Rothman \shortcite{gamache} and
spectroscopic data from the HITRAN database \cite{hitran}. 

\subsubsection{Emission features}

\begin{figure}
\setlength{\unitlength}{1cm}
\begin{center}
\leavevmode
\begin{picture}(5.0,4.)
\put(0.0,4.0){\includegraphics{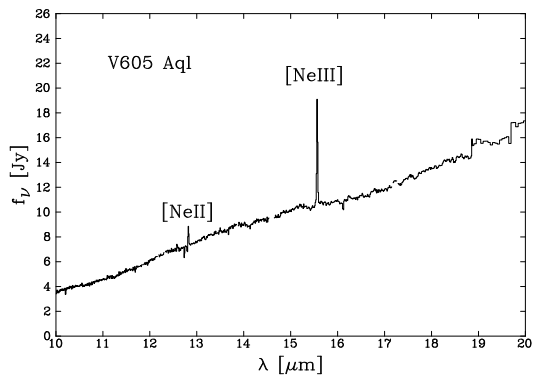}}	 
\end{picture}
\end{center}
 \caption[]{The Spitzer IRS spectrum of V605~Aql (Evans et al., in preparation).}
\label{v605}
\end{figure}

In view of the increased temperature of the stellar remnant
\cite{kerber,hajduk} we expect the appearance of IR fine structure
lines as the wind is ionized. In the case of V605~Aql, which is
$\sim100$~years ahead of Sakurai's Object in its evolution, we
clearly see \fion{Ne}{ii}~$\lambda12.8$\mic\ (line flux  
$6.0\times10^{-17}$~W~m$^{-2}$) and \fion{Ne}{iii}~$\lambda15.5$\mic\
($2.6\times10^{-16}$~W~m$^{-2}$) in the Spitzer IRS spectrum, although
the dust shell is clearly much cooler (see Fig.~\ref{v605}). However
there is no evidence for the presence of these emission lines in the
Spitzer spectrum of Sakurai's Object, to a limit of
$3\times10^{-16}$~W~m$^{-2}$ for both \fion{Ne}{ii} and \fion{Ne}{iii},
that we can attribute to increased ionization of the ejected gas as the
effective temperature of the star increases.

\section{Discussion}

\subsection{The HCN features}
\label{hcn}
The temperature of the dust ``photosphere'' is $\sim200$~K (see 
\S\ref{cont}), so we seem to have the unphysical situation that the 
temperature of the absorbing gas ($\sim450$~K) is greater than that
of the dust. Cernicharo et al. \shortcite{cernicharoetal99} note
that the 14\mic\ HCN feature can appear in emission if the HCN-bearing
material is close to the star. The fact that it is in absorption in
Sakurai's Object indicates that this material is unlikely to be close
to the star, while the kinetic temperature of 450~K indicates heating
by ultraviolet photons in a thin layer of gas. For this to occur the
radiation must penetrate the optically thick dust shell, which is possible
only if the dust distribution has strong asymmetry.

The HCN column is $\sim10^{17}$~cm$^{-2}$, but his should be
regarded as a lower limit because, as noted above, the geometry of the
dust and HCN is likely to be more complex than we have assumed here.
Indeed high resolution IR spectroscopy \cite{tyne0} around the
\pion{He}{i}$\lambda1.083$ line suggests that the distribution of 
matter in the outflow from Sakurai's Object is not spherically 
symmetric, as our {\sc dusty} modelling has assumed (again 
underlining that the dust properties and distribution are not well
known). Rather the dust may be distributed in an equatorial `torus', 
to which the He and HCN `wind' is orthogonal.
Although the Spitzer IRS does not give sufficient spectral resolution to
determine a doppler shift for HCN, in the case of the
\pion{He}{i}$\lambda1.083$ line there is clear evidence of outflow.
Tyne et al. \shortcite{tyne0} found outflows with $\sim600$\vunit.

However the deduced $^{12}$C/$^{13}$C ratio should be relatively
insensitive to the geometry.

\subsection{The H-deficient wind}

It is remarkable that the hydrogen-deficient, carbon-rich Sakurai's
Object should display absorption lines of H-containing species.
The presence of H-bearing molecules in the Spitzer IRS spectra
(Fig.~\ref{SED}{b}) might imply that these features from material
ejected by Sakurai's Object prior to its current hydrogen-deficient state.
However the $^{12}$C/$^{13}$C ratio clearly demonstrates that the
absorbing material must be a product of the VLTP.

It is therefore of interest to consider the origin of these species.
In the case of IRC+10$^\circ$216, the proto-typical carbon-rich 
post-AGB star, chemistry takes place in the photo\-chemistry zone and 
polyynes are produced via the reactions:
\[ \mbox{C}_2\mbox{H} + \mbox{C}_n\mbox{H}_2 \rightarrow \mbox{C}_{n+2}\mbox{H}_2 + \mbox{H} \]
\cite{millar}. This requires the presence of C$_2$H, which is formed
from C$_2$H$_2$ by photodissociation, i.e. polyynes do not form
before C$_2$H$_2$ is present. Furthermore,
Cernicharo \shortcite{cernicharo04} has shown, in the context of the
proto-planetary nebula CRL618, that carbon-rich molecules are formed
in the presence of a strong photo\-dissociating radiation field.

It generally seems the case therefore that the chemistry leading to
these species requires the presence of radiation. Such an environment
might exist in the case of Sakurai's Object along the `wind' axis postulated in
\S\ref{hcn}. However whether the chemistry that led to the
hydrocarbons we see in Sakurai's Object is a
relatively recent phenomenon, or whether it is the result of earlier
chemical processes whose products have been `frozen in' is unclear.

\vspace{-2mm}

\section{Conclusions}

We have reported an observation of V4334~Sgr (Sakurai's Object) with 
the Spitzer IRS. We find that the IR SED is dominated by the 
optically thick dust shell that still enshrouds the stellar remnant.
The dust shell ``photosphere'' has a temperature $\sim200$~K, while 
the temperature of the inner dust shell is inferred to be $\sim400$~K.
There is evidence that the condensation of new dust has ceased, and
that the character of the dust has changed since 2001.

There are several prominent absorption features, which we attribute 
to HCN, acetylene and polyynes. The $^{12}$C/$^{13}$C ratio is 
$\sim3$; this is substantially smaller than the Solar System value
($\sim70$) and is consistent with (a)~the CNO cycling following the
VLTP, and (b)~the value obtained from the IR first overtone CO bands.

We see no evidence for the presence of emission lines (such as
\fion{Ne}{ii}~$\lambda12.8$\mic, \fion{Ne}{iii}~$\lambda15.5$\mic)  
that would be a signature of the anticipated increased $T_{\rm eff}$
of the star.

\section*{ACKNOWLEDGMENTS}

This work is based on observations made with the Spitzer Space
Telescope, which is operated by the Jet Propulsion Laboratory,
California Institute of Technology under a contract with NASA.
Support for this work was provided by NASA through an award
issued by JPL/Caltech.

We thank Dr J. Cernicharo for comments on an earlier version of this paper.
AE thanks Dr Greg Harris, UCL, for providing essential information and
invaluable advice about the HCN transitions.
TRG is supported by the Gemini Observatory, which is
operated by the Association of Universities for Research in Astronomy,
Inc., on behalf of the international Gemini partnership of Argentina,
Australia, Brazil, Canada, Chile, the United Kingdom, and the United
States of America.
RDG, CEW and EP are supported by NASA, the NSF (AST02-05814), the US
Air Force, and the University of Minnesota Graduate School.
SGS acknowledges partial support from Spitzer and NSF grants to 
Arizona State University.
Data reduction was carried out using hardware and
software provided by PPARC.

\bsp

\label{lastpage}

\end{document}